\documentclass[sigconf,natbib=false]{acmart}
\AtBeginDocument{%
  }

\copyrightyear{2023}
\acmYear{2023}
\setcopyright{rightsretained}
\acmConference[WPES '23]{Proceedings of the 21st Workshop on Privacy in the Electronic Society}{November 26, 2023}{Copenhagen, Denmark} \acmBooktitle{Proceedings of the 21st Workshop on Privacy in the Electronic Society (WPES '23), November 26, 2023, Copenhagen, Denmark}\acmDOI{10.1145/3603216.3624966}
\acmISBN{979-8-4007-0235-8/23/11}

\usepackage{todonotes}
\usepackage{algorithm}
\usepackage{algpseudocode}
\usepackage{tabularx}
\usepackage{eurosym}
\usepackage{subcaption}
\usepackage{cleveref}
\usepackage{amsmath}
\usepackage{paralist}
\usepackage{subcaption}

\RequirePackage[
  datamodel=acmdatamodel,
  style=acmnumeric,
  ]{biblatex}

\makeatletter
\gdef\@copyrightpermission{
  \begin{minipage}{0.3\columnwidth}
   \href{https://creativecommons.org/licenses/by/4.0/}{\includegraphics[width=0.90\textwidth]{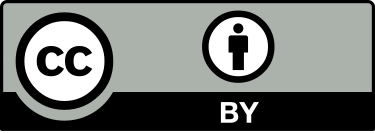}}
  \end{minipage}\hfill
  \begin{minipage}{0.7\columnwidth}
   \href{https://creativecommons.org/licenses/by/4.0/}{This work is licensed under a Creative Commons Attribution International 4.0 License.}
  \end{minipage}
  \vspace{5pt}
}
\makeatother

\begin{document}





\title{Legitimate Interest is the New Consent -- Large-Scale Measurement and Legal Compliance of IAB Europe TCF Paywalls}

\author{Victor Morel}
\affiliation{%
  \institution{Chalmers University of Technology}
  \city{Göteborg}
  \country{Sweden}}
  \email{morelv@chalmers.se}

\author{Cristiana Santos}
\affiliation{%
  \institution{Utrecht University}
  \city{Utrecht}
  \country{The Netherlands}}
\email{c.teixeirasantos@uu.nl}

\author{Viktor Fredholm}
\affiliation{%
  \institution{Chalmers University of Technology \newline Department of Computer Science and Engineering}
  \city{Göteborg}
  \country{Sweden}}
  \email{vikfre@student.chalmers.se}

  \author{Adam Thunberg}
\affiliation{%
  \institution{Chalmers University of Technology \newline Department of Computer Science and Engineering}
  \city{Göteborg}
  \country{Sweden}}
  \email{thadam@student.chalmers.se}


\renewcommand{\shortauthors}{Victor Morel, Cristiana Santos, Viktor Fredholm, \& Adam Thunberg} 
\begin{abstract}
Cookie paywalls allow visitors of a website to access its content only after they make a choice between paying a fee or accept tracking. 
European Data Protection Authorities (DPAs) recently issued guidelines and decisions on paywalls lawfulness, but it is yet unknown whether websites comply with them. 
We study in this paper the prevalence of cookie paywalls on the top one million websites using an automatic crawler. 
We identify 431 cookie paywalls, all using the Transparency and Consent Framework (TCF). 
We then analyse the data these paywalls communicate through the TCF, and in particular, the legal grounds and the purposes used to collect personal data. 
We observe that cookie paywalls extensively rely on legitimate interest legal basis systematically conflated with consent. 
We also observe a lack of correlation between the presence of paywalls and legal decisions or guidelines by DPAs.
\end{abstract}


\settopmatter{printfolios=true}
\maketitle

\section{Introduction}

Currently the Web business model employs cookie paywalls -- also named ``pay or okay''  or even simply ``paywalls'': 
if a user refuses 
tracking, she is then obliged to provide a sum of money to access that website~\cite{cnil_cookie_22}.
%
All cookie paywalls found by the literature~\cite{morel_your_2022} were using the Interactive Advertising Bureau (IAB) Europe Transparency and Consent Framework (TCF) 
-- the European-level association for the digital marketing and advertising ecosystem~\cite{iabtcf}.
Several European Data Protection Authorities (DPAs) recently ruled cases where cookie paywalls were deemed unlawful or issued guidelines imposing safeguards for these practices, for instance, the Austrian~\cite{noyb_pay_2023,austrian_dpa_dpa_2023}, Spanish~\cite{aepd_aepd_2023}, Danish~\cite{datatilsynet_cookie_2023}, and German~\cite{data_protection_lower_saxony_decision_2023,noyb_pay_2023-1} DPAs. 
However, it is yet to be determined whether websites comply with the DPA guidelines and decisions.

Thus, this paper addresses the following research questions: 1) Are IAB Europe TCF implementations of cookie paywalls in line with EU data protection law requirements? 2)  To what extent do websites follow the DPA positions concerning cookie paywalls? 
To answer these questions, we design a
crawler to detect cookie paywalls over the top one million websites, and analyse the data they communicate through the TCF.
Our contributions can be listed as follows:
(1) We measure cookie paywalls in terms of numbers, geographic distribution, categories, and prices. We found out
that cookie paywalls can now be found in several types of
websites, and mostly in Germany, despite  the critical stance of the DPA towards these.
(2) We provide an onlook at TCF implementations in the wild.
We found that websites hosting cookie paywalls 
i) rarely register consent before an action from a visitor, 
ii) extensively rely on legitimate interest legal basis for data collection, and 
iii) the TCF specifications, in spite of the V2.2 update, does not bring sufficient technical guarantees with respect to the ban of this legal basis of legitimate interest for advertising purposes.
(3) We provide an update to the legal regulatory landscape regarding cookie
paywalls. We point out that the use of legitimate interest by TCF-based paywalls as a legal basis is illegal, and that the
distribution of cookie paywalls does not seem to be affected
by national DPA decisions.

\vspace{-4mm}


\section{Background and related work}
\label{sec:background}

\subsection{Legal landscape}
\label{sec:legal}

\textbf{Consent is the lawful ground for online tracking.}
The GDPR applies to the processing of personal data ~\cite{EDPB-4-07} and requires organizations (called data controllers) to choose a legal basis  for the processing of personal data to be lawful, including the legal basis of consent and legitimate interest, amongst other grounds (Article 6(1)).
%
The ePrivacy Directive (ePD)~\cite{european_parliament_directive_2002} provides \emph{supplementary} rules to the GDPR in particular for the use of  tracking technologies. 
To comply with the GDPR and the ePD, websites must obtain \emph{consent} from EU users when tracking their  behavior (Art. 5(3) ePD)~\cite[para. 44]{edpb_-_edps_edpb-edps_2022} for concrete purposes (such as targeted advertising). 
Some other purposes are exempted of consent, \textit{e.g}, functional or essential trackers (Recital 66 ePD).
The only way to assess with certainty whether consent is required is to analyse the \emph{purpose} of each tracer on a given website~\cite{EDPB-4-12, WP203, Foua-etal-20-IWPE}.
%
A  valid consent  must comply with several requirements: \textit{prior}, \textit{freely given}, \textit{specific}, \textit{informed}, \textit{unambiguous}, \textit{readable}, \textit{accessible}, and \textit{revocable} (Art. 4(11) and 7 GDPR)~\cite{Sant-etal-20-TechReg}.
Most relevant to this paper is the requirement of a \textit{freely given consent} which means that the request for consent should imply a voluntary  choice of the user to accept or decline some or all purposes. Such choice should be taken in the absence of any kind of pressure 
 to persuade her to give consent~\cite{wp29_opinion_2011, wp29_working_2013, edpb_guidelines_2020}, or in the absence of negative consequences in case a user rejects consent to tracking. 
Making the access of a website  conditional on the acceptance  of certain non-essential trackers \textit{can affect}, in certain cases, the freedom of choice~\cite{cnil_cookie_22}, and subsequently, the validity of consent~\cite{edpb_guidelines_2020}.



\noindent
\textbf{Legal basis of legitimate interest.} Processing personal data will be lawful when it is necessary for the purposes of the legitimate interests of a data controller or by a third party to whom data was disclosed (Article 6(1)(f)).
The general provision of legitimate interest is open-ended, with a broad and unspecific scope~\cite{EDPB-6-14}, and it is not purpose-specific as long as its requirements are satisfied. 
The open-ended nature of this provision raises important questions regarding its exact scope and application~\cite{EDPB-6-14}.
It is mandatory that such processing is necessary for the purposes of a given interest.
 These legitimate interests may justify data collection if they override the data subject's interests and rights, such as the right to privacy~\cite{CJEU-FashionID, CJEU-GoogleSpain}.
Accordingly, some obligations impend over controllers: 
they are required to perform a \textit{balancing decision} in every single context as to whether this requirement is met. 

\noindent
\textbf{Legal basis for advertising purposes is consent.} 
A recent decision of the 4th July 2023 by the European Court of Justice (CJEU) in Meta vs Bundeskartellamt Case C-252/21~\cite{ecj_case_2021} established that i) consent is the appropriate legal basis for the tracking-and-profiling-driven personalized content and behavioral advertising, and ii) no legitimate interest  would override the users' rights when websites try to provide ads (see notably paragraph 117).
%

\noindent
\textbf{Paywalls are endorsed by the European Court of Justice.}
Paragraph 150 of the decision~\cite{ecj_case_2021} permits paywalls if a given fee is necessary and appropriate.
%
It then falls upon websites to motivate and inform users about a fee necessity and its appropriateness.

\noindent
\textbf{Interaction between consent and other lawful grounds.} 
%
%
Following \textit{noyb}'s~\footnote{\textit{noyb}, which stands for ``not of your business'' is a ``European non-profit using strategic litigation to enforce [the] fundamental right to data protection and privacy''.} cognition posited in its complaints, if a controller requests user consent, this choice has a \emph{blocking effect} regarding other legal basis, \emph{i.e.} the website has deprived itself of the possibility of basing the data processing on another legal basis under Article 6(1) of the GDPR.~\footnote{See Section 3.4.1 of the various complaints that were issued by \textit{noyb}~\cite{noyb_news_2021}.}
%
%
%
Moreover, considering the understanding of the European Data Protection Board guidelines~\cite[parag 121-123]{edpb_guidelines_2020}, the application of one of the six legal bases under Article 6(1) must be disclosed \emph{prior} to data collection and in relation to a specific purpose. This means that if a controller chooses to rely on consent for any part of the processing,  while actually another lawful basis is relied upon, this would be fundamentally unfair to individuals. Thus the controller cannot swap from consent to other lawful bases, like legitimate interest. 
%
%




\noindent\textbf{IAB Europe Transparency and Consent Framework.} 
IAB Europe TCF defines in its specifications ten purposes that can rely on both legal grounds: consent and legitimate interest.
The purposes are: 
\begin{inparaenum}
    \item Store and/or access information on a device,
    \item Select basic ads,
    \item Create a personalised ads profile,
    \item Select personalised ads,
    \item Create a personalised content profile,
    \item Select personalised content,
    \item Measure ad performance,
    \item Measure content performance,
    \item Apply market research to generate audience insights, and
    \item Develop and improve products.
\end{inparaenum}
In February 2022, this framework was declared to infringe the GDPR for using unlawful practices and for collecting data for advertising purposes on the ground of legitimate interests~\cite{APD2022,Veale_Nouwens_Santos_2022}. 
In September 2022, the TCF was brought to the highest court of the EU (the European Court of Justice)~\cite{belgian_dpa_iab_2022}.
In order to comply with the law, IAB Europe recently announced the new version 2.2 of the TCF (to be enforced in September 2023), which will notably prevent the use of legitimate interest for purposes 3, 4, 5, and 6~\cite{tcf_tcf_2023}.
%


\vspace{-2mm} 

\subsection{Related work}
\label{sec:related}
\textcite{papadopoulos_keeping_2020} automated the detection and classification of paywalls on the Web involving machine learning, but the study did not address \textit{cookie} paywalls specifically.
The authors did however conduct a thorough review of the types of sites that employ paywalls and their country of origin. 
%
Matte et al.~\cite{matte_cookie_2020} investigated 28 257 websites, of which 1 426 implemented the TCF.
They found that more than 50\% of the web pages analysed did not comply with the GDPR or the ePD, and that all non-compliant 
web pages implemented the TCF.
The only technical and legal study on cookie paywalls was conducted by \textcite{morel_your_2022}.
The study was based on a manual classification of the most popular websites in 13 Central European countries. 
They analyzed 2800 websites and found 13 websites employing cookie paywalls.
They used a heuristic method based on features of the language of cookie paywalls to detect them. 
The 13 cookie paywall websites were analysed to extract data such as the type of banner (e.g. blocking or not), website category, and the price/type of subscription. 
 They provided a legal analysis of cookie paywalls in the light of EU data protection law and regulatory guidelines, and also presented a fine-grained classification of paywalls.
We have built upon the work of this study by 1) performing similar analyses on a larger scale and programmatically,
and 2) providing an update to the legal landscape with regulatory decisions and recent DPA guidelines.


\vspace{-2mm} 

\section{Methodology}
\label{sec:crawling}
We present here how we designed our cookie paywall crawler, and how we analysed the data they convey through the TCF.

\noindent
\textbf{Crawler.}
\label{sec:crawler}
We built a crawler to identify cookie paywalls using text processing on the top 1 million URLs using the Daily List from Tranco~\cite{le_pochat_tranco_2019}.~\footnote{\url{https://tranco-list.eu/list/PZPLJ/1000000}.}
%
%
The crawler was configured to run 32 agents using a matching 32 Firefox browsers. 
Each of the agents and browsers ran in their own containers inside of a Kubernetes cluster. The container images for the browsers were pulled from the Docker Selenium project,
which in turn makes use of the Firefox version 113.0.1.
Once the crawl was completed, cookie paywall sites flagged as ``likely'' were manually confirmed by two independent annotators,
along with a classification of geographical basis, website category, and paywall price per month. 
%
The country in which the site is based was determined by analyzing the WHOIS requests for each site.~\footnote{WHOIS is a protocol used to determine the registered owner of an internet domain name. 
Although the quantity and type of information can be inconsistent when querying a WHOIS database, it often contains geographical information associated with the registrant of a domain (typically a company in the case of cookie paywall websites).}
We developed a script that takes the response of a WHOIS query of a domain name and looks for fields such as ``country:''. 
If no relevant fields was found, we used the domain TLD (such as .de for Germany).
%
The website type was determined by Cyren's URL Category Checker~\cite{cyren_url_category_check}. 
They use a URL classifier to assess threats from web pages, provided on their website~\cite{cyren_website_nodate}.
The classifier sometimes returned two categories for the same URL.
In that case, only the first category was considered as we assume it is the most likely match (following Cyren's guidelines).

\noindent
\textbf{TCF analysis.}
\label{sec:tracking}
From the results of the crawl, two sets of cookie paywalls were distinguished: i) one set containing cookie paywalls using the Consent Management Platform (CMP) \textit{contentpass}~\footnote{Although \textit{contentpass} can be considered a Subscription Management Platform (SMP).} which exclusively provides cookie paywalls to 220 websites (189 of which were analysed); and ii) another set using other various CMPs.
We then used three different approaches as explained below.
In all approaches, we extracted the number of vendors to which data is conveyed based on both consent and legitimate interest.
%
The data resulting from the analysis can be found following this \href{https://github.com/Norrland97/cookie_paywall_crawl}{link}.
%
%
%
%
In the \noindent\textit{automated approach}
\label{Section:contentpass}
    a list of all websites implementing \textit{contentpass} was created by scraping its marketing webpage.
    Every website in the list was analysed, and the TCF consent string was stored as it appears in all three relevant states -- before interaction, after giving consent, and after logging in.
    The TCF consent string was extracted by searching the cookie jar and local storage for cookies stored per the naming standards specified in the TCF.
\label{section:semi-auto}
We retrieved data from a large and varied set of cookie paywalls using a \noindent\textit{semi-automated approach} to 
get an overview of how the cookie paywalls behave before any interaction with a visitor.
We browsed all non-\textit{contentpass} websites with a script which automatically saves the consent string recorded by the website.
    %
In our \noindent\textit{manual approach}
\label{Subsection:Manually}
we performed a manual analysis on a subset of websites \textit{not} implementing \textit{contentpass}, for which we paid a subscription.
We randomly selected 20 websites for manual inspection using pythons built-in random generator. 
%
In the case where one subscription was giving access to several cookie paywalls,
we investigated whether the implemented cookie paywalls -- which are part of the same subscription --, differ from each other.


\vspace{-2mm} 

\section{Results and discussion}
\label{sec:results}

%

\noindent \textbf{All found cookie paywalls use the controversial TCF}.
108 cookie paywalls were found in a preliminary calibration phase.
The crawler then processed 1 million pages in about 5 days, reporting 330 as ''likely'' (see Section~\ref{sec:crawler}).
The confirmed number was 323, giving the crawler a 
positive accuracy of \textasciitilde{}98\%. 
All confirmed paywalls, along with their assigned classification and price, can be found in this \href{https://docs.google.com/spreadsheets/d/1UBiIaH5LAf04IlDsnf7zm68b_csA4_KiJDZyaZcLdBc/edit}{Google Sheet}.
%
%
They were combined with the 108 initially found in the preliminary phase, making a total of 431, all using the TCF.
Note that the lawfulness of this framework is currently being argued at the Court of Justice of the EU (IAB Europe (C-604/22))~\cite{brussels_markets_court_iab_2022}
and the TCF was considered illegal by the Belgian DPA~\cite{APD2022, Veale_Nouwens_Santos_2022}. 

\begin{figure*}
\hspace*{-2cm}
    \begin{subfigure}[b]{.6\textwidth}
\centering
    \includegraphics[width=\textwidth]{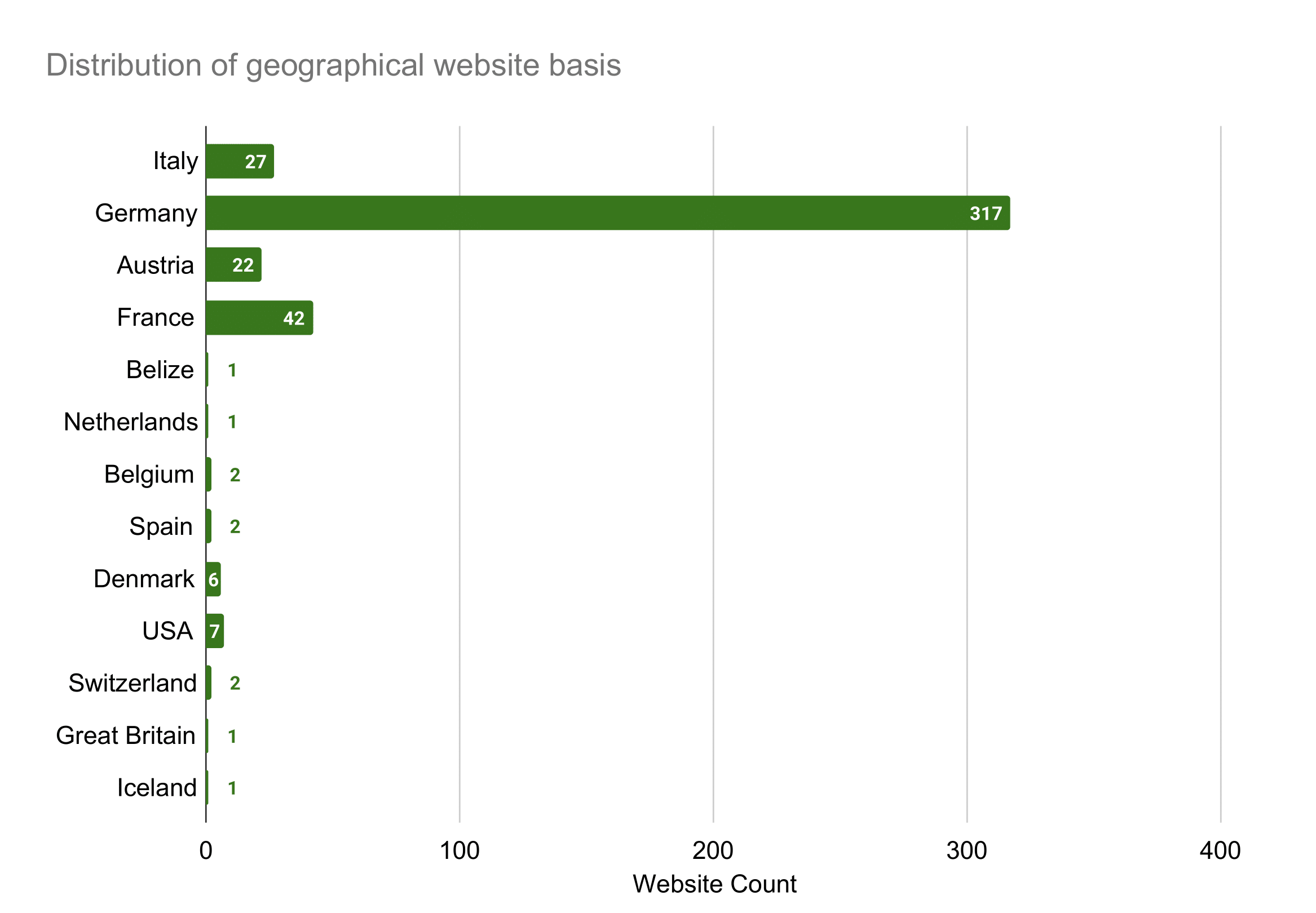}
    \caption{Distribution of cookie paywalls found per country}
    \label{fig:dist-countries}
    \end{subfigure}
    ~
    \begin{subfigure}[b]{.6\textwidth}
    \centering
    \includegraphics[width=\textwidth]{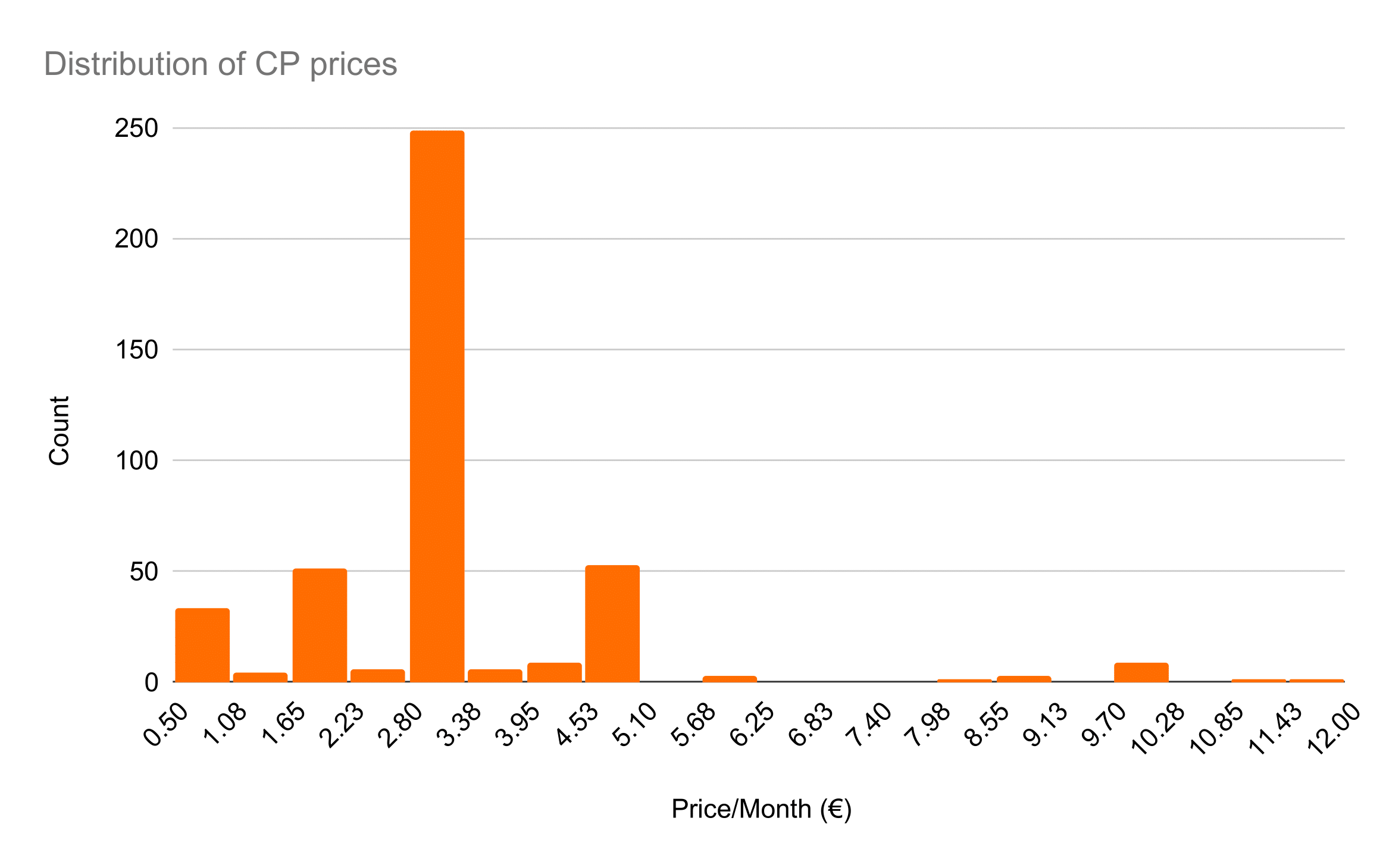}
    \caption{Distribution of cookie paywall prices in euros per month}
    \label{fig:prices-chart}
    \end{subfigure}  
    \caption{Distribution of cookie paywalls}
\end{figure*}

\noindent
\textbf{Paywalls are prevalent in Germany despite the DPA stance.}
The distribution of cookie paywalls across all countries is depicted in Figure~\ref{fig:dist-countries}. 
Our results show a preponderance of cookie paywalls in Germany (317 out of 431), followed by France (42), Italy (27), and Austria (22) -- the other countries having only a few cookie paywalls (between 1 and 6). 
However, the position of the German DPA is critical of cookie paywalls~\cite{morel_your_2022}.
This fact indicates that in Germany the prohibition of cookie paywalls may not  affect their prevalence. The DPAs of other countries prefer to assess paywalls case by case, as shown in Table~\ref{table:requirements}. 
%
One could reason that the existence of the German-based CMP \textit{contentpass} might justify the concentration of paywalls in Germany.
This particular CMP only offers a cookie paywall solution, and it doesn't provide other type of cookie banners. 
However, in a closer look, while 317 cookie paywalls were found in Germany, only 220 used \textit{contentpass} which means that 97 are \textit{not} using it. 
Such number is  considerably higher than the runners-up France, Italy, and Austria, which indicates that \textit{contentpass} is not  the only reason for paywall prevalence in Germany.

\noindent
\textbf{Users consent when facing \textit{contentpass} paywall but they are tracked by up to 365 adtech vendors.}
When contacting the CEO of \textit{contentpass} to better understand their solution,
we were informed that 99.9\% of visitors consent when facing a \textit{contentpass} paywall,~\footnote{At the time of publication of the present paper.} and therefore do not pay (in spite of the first month of subscription being free). 
This fact -- although not a direct result from our crawling -- indicates that websites using \textit{contentpass} do not rely on subscription but rather on ad revenues for their business model.
 Notably, after giving consent, when decoding the consent string of a website using \textit{contentpass}, 
for instance \href{https://iabtcf.com/#/decode?tcstring=CPvxKXAPvxKXAAfKGBENDQCsAP_AAH_AAAYgJStd_H_fbX9j-f59aft0eY1f9_L_buQyDheFo-oFyNeQ9LwG22E6NEygpCgCkR4golJBIANsHElcCUERQAgFAAHsCgAEpAAIICBEgBMZQkIICBoKAoQQQACJgEg9MhWImwqWY9LmXEEA9IgQBggAgIAgAAIBAgMARAAIABIAAQIAgYAAAAAAMEIAAAACABgAAAQAAAAAIJStd_H_fbX9j-f59aft0eY1f9_L_buQyDheFo-oFyNeQ9LwG22E6NEygpCgCkR4golJBIANsHElcCUERQAgFAAHsCgAEpAAIICBEgBMZQkIICBoKAoQQQACJgEg9MhWImwqWY9LmXEEA9IgQBggAgIAgAAIBAgMARAAIABIAAQIAgYAAAAAAMEIAAAACABgAAAQAAAAAIAA}{https://www.spielfilm.de/}, personal data is 
shared with up to 365 vendors. These vendors include major adtech vendors and data brokers such as Oracle Advertising, Criteo SA, and Acxiom.
We question whether consent is freely given -- even if an alternative to tracking exists (i.e. a subscription) -- since personal data is being shared with so many third parties, which might render tracking detrimental. 

\begin{table}[!ht]
\small\addtolength{\tabcolsep}{-2pt}
\footnotesize
\begin{tabular}{p{1.8cm}|p{6cm}} 
    \hline
    \textbf{DPAs} & \textbf{Positioning on cookie paywalls}\\
	\hline
German DPA~\cite{data_protection_lower_saxony_decision_2023} &  Recent case in which ``Pay or Okay'' approach was ruled illegal for an online newspaper
 \\
\hline
Spanish DPA~\cite{aepd_aepd_2023} & Guidelines state that access cannot be conditioned to consent to cookies, but exceptions can be made if alternatives are offered (not necessarily free ones) and users informed
\\
	\hline
	French DPA~\cite{cnil_cookie_22} & Case by case assessment.  Websites need to show there is a real and fair alternative way to access other websites without tracking; reasonable price; fair remuneration\\
\hline
	    Austrian DPA~\cite{austrian_dpa_dpa_2023} & Dual position: Recent decision: paywalls are generally permissible, but users must have the possibility to say "yes" or "no" to any specific data processing.    \\ \hline
\end{tabular}
    \caption{DPAs positioning regarding cookie paywalls.} 
    \label{table:requirements}
    \vspace{-8mm}
\end{table}

\begin{figure*}
\centering
    \includegraphics[width=\textwidth]{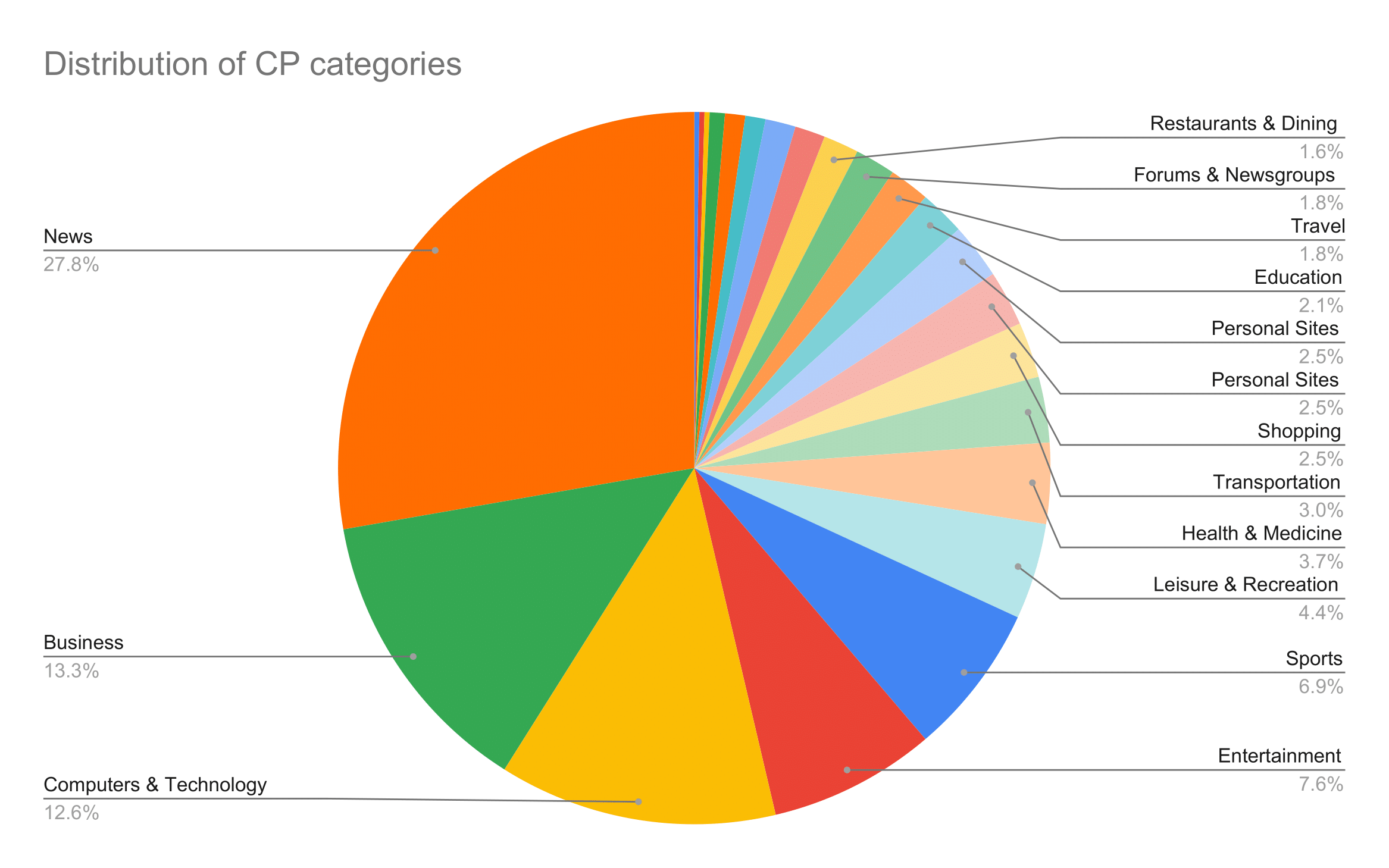}
    \caption{Distribution of website categories hosting cookie paywalls, as reported by Cyren~\cite{cyren_url_category_check}.}
    \label{fig:dist-categories}
\end{figure*}

\noindent
\textbf{Paywalls are not restricted to news any longer, they are spread into business, tech, and entertainment websites.}
The distribution of website categories (see Figure~\ref{fig:dist-categories}) shows that a large number of the found cookie paywalls were classified as News (27.4\%), confirming former work, as paywalls improve content monetisation and thus fund journalism~\cite{euractiv_austria_2023}.
%
The frequency of  paywalls on sites in the Business (13.2\%), Computer \& Technology (12.3\%), and Entertainment (7.7\%) categories suggest a potential reliance on a combination of subscription revenue and sharing of personal data in these sectors as well.
These categories include sites that often host high-traffic platforms that attract a large user base that can be leveraged for targeted advertising purposes.

\noindent
\textbf{Paywalls seem to have a reasonable cost - €3.34 on average.}
All cookie paywalls used a monthly subscription-based payment model, wherein the vast majority (67\%) cost between €2 and €4 per month,  with an average price of €3.34 per month. 
The distribution of price is visualized in Figure~\ref{fig:prices-chart}.~\footnote{Only 3 sites had a price above €10, with one notable site with a monthly price of €49 per month left out of the chart for readability.}
%
\textcite{morel_your_2022} argued that according to the French DPA, 
the cost of a paywall should be ``reasonable'' or consist of a ``fair remuneration''~\cite{cnil_cookie_22}.
As mentioned on Section~\ref{sec:legal}, neither the CJEU established what is a necessary or appropriate fee, and so the determination of prices is yet to be further studied and harmonized.
It is worth noting that \textit{contentpass} offers a cross-site subscription-based model of €2.99 per month.
This finding can be read in the light of Mueller-Tribbensee et al. study's results, in which the authors argued that 99\% of users tend to consent when facing paywalls.
We thus conclude that even if the price seems reasonable, users choose to be tracked~\cite{data_protection_law_scholar_network_dpsn_2023}.

\noindent
\textbf{The TCF conflates the legal grounds of consent and legitimate interest.}
We observed that \textit{all} websites hosting cookie paywalls systematically communicate consent strings registering purposes under 
legitimate interest -- in addition to consent -- after a visitor clicks on ``accept''.~\footnote{Note that not all purposes are not necessarily communicated under legitimate interest.}
As commented in Section \ref{sec:legal}, if a website requests consent, her choice has a \emph{blocking effect} regarding other legal basis,  such as legitimate interest, and thus the website is deprived of  processing on another legal basis~\cite{noyb_news_2021}. We argue that such overlapping renders processing illegal. 


\noindent
\textbf{Even if users pay for a subscription, they are still tracked under legitimate interest for inappropriate purposes.} 
    After a paid subscription to paywalls, some websites (14 websites, including 13 using \textit{contentpass}) still collect  personal data based on legitimate interest \emph{by default}. 
   This means that users have to manually object if they wish to avoid their data to be collected. Firstly, users should not be anyway tracked if they pay for a subscription, since paywalls are only legitimate if they consist of an alternative to tracking.  
  Secondly,  3 of these websites track users for the purpose to ``\textit{Develop and improve products}'' under the legal ground of legitimate interest. This purpose is vague and unspecified, since it is not detailed enough to determine its kind of processing~\cite{WP203,edpb_guidelines_2020} and therefore illegal ~\cite{matte_purposes_2020}. 
  The European Data Protection Board  (EDPB) guidance~\cite {EDPB-guidelines19}  proposes legitimate interest  for this purpose under  detailed information on how users engage with their service through a organizational metrics for a concrete service, also grounding the way to improve it, and cannot be used in general, as the TCF uses.
  Thirdly, 3 websites collect data for 5 purposes under legitimate interest including ``\textit{Select basic ads}'', ``\textit{Measure ad performance}'', ``\textit{Measure content performance}'', ``\textit{Apply market research to generate audience insights}'', and ``\textit{Develop and improve products}'' (\url{https://karrierefragen.de} for instance). 
  However, consent is always required for advertising related purposes~\cite{ecj_case_2021}, and for third-party analytics~\cite[p~.47]{EDPB-6-14}. We take the view that the latter  purpose \emph{``Apply market research to generate audience insights''} is defined in a broad way and with ambiguity as to its intent ~\cite{matte_purposes_2020}.

%


\noindent
\textbf{Although the TCF will disable the use of legitimate interest for certain ad-related purposes, websites can design custom storage of advertising purposes under the legal basis of legitimate interest.}
\label{subsection:custom}
When looking for how consent is stored in the local storage of the browser, we observed that
some purposes reliant upon the legal basis of legitimate interest were found to be stored separately from the TCF consent string 
%
-- under the string \textit{gdpr -> customVendorsResponse -> legIntPurposes} (e.g. \url{www.voici.fr}).
Some of these identified purposes are ad-related (purposes 3, 4, 5, and 6, see Section~\ref{sec:background}),
but according to the legal background laid down in Section~\ref{sec:legal} should not rely under this legal basis. 
The legitimate interest related purposes  were found in 12 of the manually inspected websites. 
11 of these websites have the same CMP (Prisma Media), so this customisation may only be permitted by certain CMPs.
This finding needs to be interpreted in the light of the TCF v2.2 update which will purportedly prevent the use of legitimate interest for personalised advertising~\cite{tcf_tcf_2023}.
Indeed, this storage customisation of user choices i) may hinder the monitoring of the TCF framework by IAB, ii) legitimize non compliance practices from CMPs, and iii) may enable the circumvention of the technical safeguards brought by the upcoming TCF update.

\noindent
\textbf{Limitations.}
The majority of cookie paywalls detected in our study are European-based, only 8 were found outside thereof.
This may be due to a biased methodology in that our detection algorithm was based on a previous study that specifically focused on identifying European cookie paywalls. 
To gain a deeper understanding, future research should examine different geographic regions and develop detection algorithms that are less specific to certain regions.

\vspace{-2mm}

\section{Conclusion and recommendations}
\label{sec:conclusion}
We 
found 431 cookie paywalls and reported that 
most paywalls were found in Germany despite of its DPA positioning.
%
These paywalls extensively use advertising purposes under the legal basis of legitimate interest 
to collect personal data. 
Promising research points to 1) 
a browser extension to bypass paywalls, and 2) the assessment of paywalls on mobile browsers.
%
Based on our findings, we also formulate policy recommendations.
\textbf{First}, because we observed that consent strings can be stored on local storage (as opposed to regular cookies), we recommend the ban of custom storage for legitimate interest-based purposes, as it can include  advertising (see Section~\ref{subsection:custom}) and renders auditability difficult.
\textbf{Second}, since the TCF consent string communicates purposes for both consent and legitimate interest, we advocate for a compliant signal which clearly distinguishes the two legal grounds.
\textbf{Third}, considering that isolated DPA decisions and their guidelines may not be enough to bound websites, and that the CJEU decision is yet not clear about what is a necessary and appropriate fee, we call for a concerted and harmonized effort from the EDPB to issue guidelines ascertaining the lawfulness of paywalls. 

\newpage

\section*{Acknowledgments}
This work was partially supported by the Wallenberg AI, Autonomous Systems and Software Program (WASP) funded by the Knut and Alice Wallenberg Foundation, as well as OmegaPoint AB.
We would like to thank Felix Mikolasch for providing us with helpful feedback for our paper.



\printbibliography

\appendix

\section{Figures}


\end{document}